\documentstyle[epsfig,longtable]{aipproc}

\begin{document}
\title{Probing the Geometry of Supernovae with Spectropolarimetry}

\author{Douglas C. Leonard, Alexei V. Filippenko, and Thomas Matheson}
\address{Department of Astronomy, University of California at Berkeley,
Berkeley, CA 94720-3411}
\maketitle

\begin{abstract}
We present results from a spectropolarimetric survey of young supernovae
completed at the Keck Observatory, including at least one example from each of
the major supernova types: Ia (1997dt), Ib (1998T, 1997dq), Ib/c-pec (1997ef),
IIn (1997eg), and II-P (1997ds).  All objects show evidence for intrinsic
polarization, suggesting that asphericity may be a common feature in young
supernova atmospheres.
\end{abstract}

\section*{Introduction}

Are supernovae (SNe) round?  This simple question belies a menacing
observational challenge, since all extragalactic SNe remain unresolvable point
sources throughout the crucial early phases of their evolution.  Since a hot
young supernova (SN) atmosphere is dominated by electron scattering, which, by
its nature, is highly polarizing, a powerful tool for investigating SN geometry
is spectropolarimetry of the expanding fireball shortly after the explosion
(Fig. 1).

\begin{figure}[hb!]
\begin{center}

 \scalebox{0.3}{
	\includegraphics{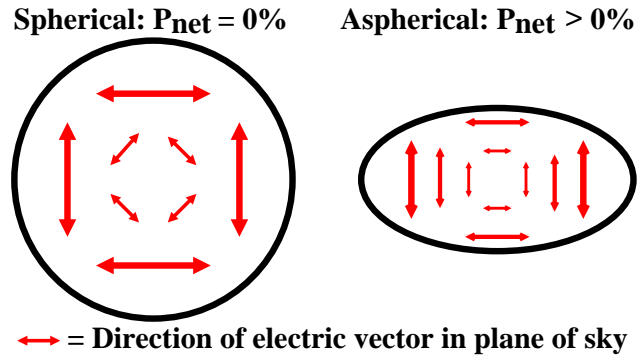}
		}
\end{center}
\caption{Polarization magnitude and direction in the plane of the sky for
resolved electron-scattering atmospheres; for an unresolved source (i.e., a
SN), only the {\it net} magnitude and direction can be measured.  An aspherical
source produces a net linear polarization due to incomplete cancellation of the
directional components of the electric vectors.  Note that the more highly
polarized light (longer arrows) comes from the limb regions in this simple
model.}
\label{fig1}
\end{figure}

Typical polarizations of $\sim 1\%$ are expected for moderate ($\sim 20\%$) SN
asphericity \cite{hoflich91}.  Detecting such low polarization requires a very
high signal-to-noise ratio, which has limited previous detailed
spectropolarimetric studies to only the two brightest recent events, SN 1987A
\cite{jeffery91} and SN 1993J \cite{trammell93,tran97}.  We thus began a
program to obtain spectropolarimetry of nearby SNe using the 10-m Keck
telescopes.

A complication in the interpretation of all polarization measurements is
disentangling the polarization intrinsic to the object from interstellar
polarization (ISP) produced by dust along the line-of-sight.  Fortunately, the
ISP is constant with time and a smoothly varying function of wavelength.
Therefore, we consider distinct spectral polarization features, temporal
changes in the overall polarization level, or continuum polarization
characteristics differing from the known form produced by interstellar dust as
evidence for intrinsic SN polarization.

\begin{figure}[b!] 
\centerline{\epsfig{file=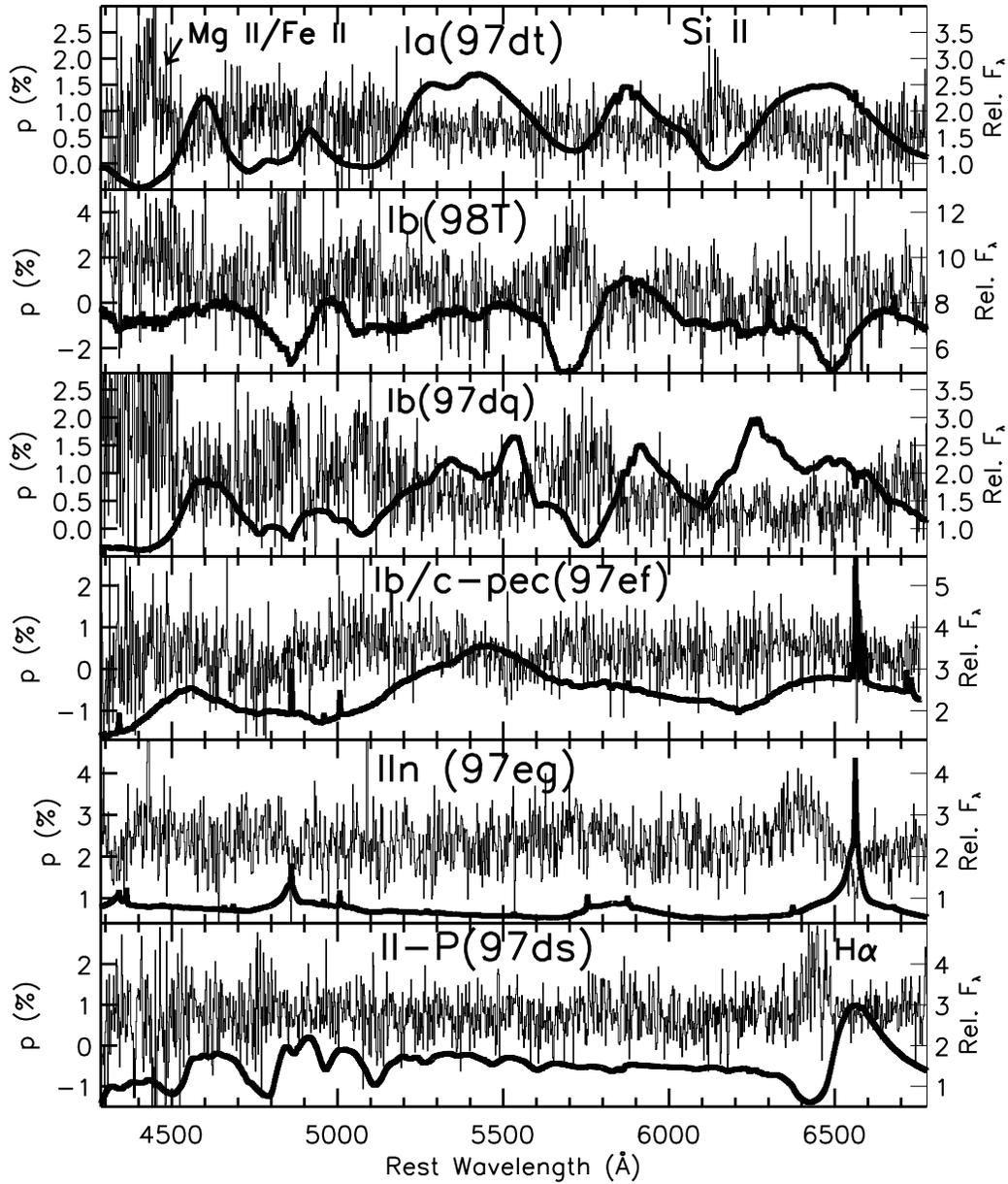,height=6.6in,width=5.0in}}
\vspace{10pt}
\caption{Spectropolarimetry (thin lines) and total flux spectra (thick lines)
of nearby SNe, all within two months of the explosion.  The recession
velocities of the host galaxies have been removed for all objects.  Note that
arbitrary, but allowable (from reddening considerations), estimates of the ISP
have been removed from the polarizations of SNe 1997dt, 1998T, 1997dq, and
1997ds in order to better highlight the line features.  Since the ISP is in
general not known, the intrinsic continuum polarization levels are uncertain;
spectropolarimetric line features, however, exist regardless of the ISP,
indicating polarization intrinsic to all of these SNe.}
\label{fig2}
\end{figure}

\section*{Results and Discussion}

Single-epoch polarization data for six SNe of various types are shown in
Fig. 2.  Since determining the intrinsic SN polarization level requires
knowledge of the (unknown) ISP, we focus instead on the sharp changes seen in
the polarization at the location of strong features in the total flux spectra;
these features remain, regardless of the ISP contribution.  Since all the
objects studied possess spectropolarimetric line features, we conclude that
{\it all} types of SNe show evidence for intrinsic polarization at early times,
suggesting that asphericity may be a ubiquitous SN characteristic.

The fact that the strongest spectropolarimetric features are often seen in the
troughs of strong P-Cygni lines is not surprising.  A simple explanation may be
that P-Cygni absorption selectively blocks photons coming from the central,
more forward-scattered (and thus less polarized) regions, thereby enhancing the
relative contribution of the more highly polarized photons from the limb
regions (c.f., Fig. 1).  Unfortunately, since different (allowable) choices for
the ISP can make inferred intrinsic polarization dips become peaks and
vice-versa \cite{leonard99a}, we cannot say for certain whether the changes
seen here in P-Cygni troughs represent increases or decreases in the intrinsic
polarization level.  We do note, however, that trough polarization increases
are seen in the ISP-corrected data of both SN 1987A \cite{jeffery91} and SN
1993J \cite{tran97}.

A total flux spectrum dominated by strong line emission without P-Cygni
absorption is a distinguishing characteristic of SNe IIn \cite{schlegel90},
likely resulting from an intense interaction between the SN and a dense
circumstellar environment (CSM).  SN 1997eg (Fig. 2) shows sharp polarization
changes across its strong, multi-component emission lines, suggesting distinct
scattering origins for the intermediate (full width at half maximum (FWHM)
$\approx$ 2000 km/s) and broad (FWHM $\approx$ 15000 km/s) components.  Two
additional spectropolarimetric epochs (not shown) revealed a change in
continuum polarization level of $\sim 1\%$ over 78 days, further confirming the
presence of intrinsic polarization.  A detailed analysis combining
spectropolarimetry and total flux spectra of another IIn event, SN 1998S, also
found evidence for a highly aspherical ($\gtrsim 45\%$) continuum scattering
region, with the CSM likely distributed in a disk-like or ring-like morphology,
quite similar to what is seen directly in SN 1987A \cite{crotts99}, except much
closer to the progenitor in the case of 1998S \cite{leonard99a}.

\section*{Conclusion}

The number of SNe studied spectropolarimetrically is still very small, but
early indications are that all types reveal intrinsic polarization if examined
in sufficient detail.  In addition to the implications of spectropolarimetry on
the core-collapse mechanism, the mass-loss history of evolved stars, and the
spatial distribution of SN ejecta, this work has direct consequences on the use
of SNe as cosmological distance indicators.  Although the empirically-based,
standard-candle technique used to measure SN Ia distances does not rely on
spherical symmetry, distances derived to SNe II-P through the ``expanding
photosphere method'' \cite{kirshner74} would need to be corrected for
directionally-dependent flux if asphericity is found to be common among this SN
class.  We note that SN 1999em, a type II-P event discovered shortly after this
conference, showed no evidence for intrinsic polarization when it was observed
less than two weeks after the explosion \cite{leonard99b}.  It will be
interesting to see if it remains unpolarized at an age comparable to the II-P
observation (SN 1997ds, observed $\sim 50$ days after explosion) presented here.

\section*{Acknowledgments}

We thank Aaron Barth for useful discussions and assistance with the
observations and data reduction.  Supernova research is supported at UC
Berkeley through NSF grant AST-9417213 and NASA grant GO-7434.


\end{document}